# Analysis of Deep Complex-Valued Convolutional Neural Networks for MRI Reconstruction

Elizabeth Cole[1], Joseph Cheng[1], John Pauly[1], Shreyas Vasanawala[2]
[1]Department of Electrical Engineering, Stanford University
[2]Department of Radiology, Stanford University

*Abstract*–Many real-world signal sources are complex-valued, having real and imaginary components. However, the vast majority of existing deep learning platforms and network architectures do not support the use of complex-valued data. MRI data is inherently complex-valued, so existing approaches discard the richer algebraic structure of the complex data. In this work, we investigate end-to-end complex-valued convolutional neural networks - specifically, for image reconstruction in lieu of two-channel real-valued networks. We apply this to magnetic resonance imaging reconstruction for the purpose of accelerating scan times and determine the performance of various promising complex-valued activation functions. We find that complex-valued CNNs with complex-valued convolutions provide superior reconstructions compared to real-valued convolutions with the same number of trainable parameters, over a variety of network architectures and datasets.

*Index Terms*— MRI, image reconstruction, complex-valued models, learning representations, convolutional neural networks

## I. Introduction

MAGNETIC resonance imaging (MRI) is a useful medical imaging technique which, unlike computed tomography, does not use harmful ionizing radiation. However, this imaging modality is relatively slow. A typical scan requires that patients remain still for a long period of time to produce images of diagnostic quality. MRI scan times can be significantly reduced by undersampling k-space at sub-Nyquist rates in various sampling patterns.

Traditionally, reconstructing images from these accelerated scans has involved leveraging techniques such parallel imaging [1], [2] and compressed sensing (CS) [3]. More recently, convolutional neural networks (CNNs) have been shown to provide a rapid and robust solution to MRI reconstruction as an alternative to slow iterative solvers. These reconstruction networks span a vast range of architectures and techniques. Examples include variational networks [4], [5], generative adversarial networks [6], [7], ADMM-Net [8], MoDL [9], unrolled methods [10], [11], hybrid networks [12], [13], U-Nets [14], [15], and AUTOMAP [16]. Deep neural networks have been quite powerful in various image reconstruction problems [4], [8], [10], [17]. Some work has even explored semi-supervised and unsupervised image reconstruction which need less or even no fully-sampled data for training [18]–[21].

While many of these networks may provide high quality reconstructions, two limitations exist with the current state of CNNs for MRI reconstruction. First, these CNNs contain millions of trainable parameters. Therefore, these networks take a long time to train and they occupy a large amount of memory, which is one of many obstacles to the practical deployment of deep learning models. Second, the vast majority of the current deep learning frameworks do not support complex-valued deep learning, even though MRI data is complex-valued. Therefore, most reconstruction networks split the real and imaginary components into two separate real-valued channels [4]–[6], [11]–[13], [16]. However, doing this discards some of the complex algebraic structure of the data and does not necessarily maintain the phase information of the data as it moves throughout the network.

In many signal processing domains, including but not limited to MRI, the handling of complex numbers is essential. However, complex representations have not traditionally appeared in many deep learning architectures because most standard computer vision datasets are real-valued. In MRI, the signals collected are complex-valued with both a real and imaginary component. The practice of separating the real and imaginary components into two channels may originate from how the colors of RGB images were split into three channels. However, splitting real and imaginary components into channels may not be the best way to represent this data structure. Recent work has shown the representative power and accuracy of complex-valued deep neural networks with applications to speech spectrum prediction and music transcription, as shown in [22], which motivates the application of complex-valued CNNs to MRI reconstruction.

The motivation for using complex numbers in convolutional neural networks stems from computational and signal

Elizabeth Cole (e-mail: ekcole@stanford.edu) is with the Department of Electrical Engineering, Stanford University, CA 94301, USA.
Joseph Cheng (email: jycheng@alumni.stanford.org) is with Apple Inc., Cupertino, CA 95014, USA; this work was done while at Stanford University.
John Pauly (email: pauly@stanford.edu) is with the Department of Electrical Engineering, Stanford University, CA 94301, USA.

Shreyas Vasanawala (email: vasanawala@stanford.edu) is with the Department of Radiology, Stanford University, CA 94301, USA.
This work was supported by NIH R01-EB009690, NIH R01-EB026136, and GE Healthcare.

processing perspectives. In the computational domain, complex weights have been shown to be more stable and numerically efficient in memory access architecture [23]. Complex gated recurrent cells, developed for recurrent neural networks, have exhibited good stability and convergence as well as competitive performance on two tasks: the memory problem and the adding problem [24], [25]. In a signal processing context, the use of complex numbers enables the accurate representation of both magnitude and phase, which are two essential components of certain signals. For example, in human speech recognition, a large amount of error in the phase of speech signals has been shown to affect the speech recognition accuracy [26]. Phase is also valuable in many MRI applications including blood flow, quantitative susceptibility mapping (QSM), fat-water separation, chemical shift imaging, and brain segmentation. Thus, using complex numbers to construct a network which more accurately reconstructs the phase is very likely to improve various MRI applications.

Several reasons could potentially explain a possible performance increase in complex-valued networks over real-valued networks for MRI applications. First, maintaining phase information throughout the network is important when the data is complex-valued because many MRI applications such as quantitative susceptibility mapping, 4D flow, and phase-contrast imaging use the reconstructed phase as valuable information. The phase is delinked in a two-channel real and imaginary convolutional neural network because different weights are applied to both input channels, altering the phase. Second, by using complex-valued convolution, the number of parameters in each model is decreased [22]. This decreases the memory a network occupies, the number of learnable parameters, and the training time. Finally, the complex-valued weights have been shown to contain higher representative power compared to real-valued weights. For example, complex numbers have been lauded for enabling faster training [33], showing smaller generalization error [34], and even allowing the network to have richer representational power [33], [35].

A few deep learning networks applied to both quantitative MRI and MRI reconstruction have demonstrated that deep learning performance can be improved over real-valued networks by using complex-valued networks [27]–[31]. Reference [28] proposed various complex-valued activation functions, such as the siglog and the complex cardioid, for solving the MRI fingerprinting inverse problem. An analogous work also proposed learning a complex-valued activation function for MRI fingerprinting CNNs [27]. Reference [29] applied a complex dense convolutional network with a U-Net based architecture to MRI reconstruction by using complex convolution and complex batch normalization. Reference [31] applied a complex dense convolutional network with a U-Net based architecture to MRI reconstruction by using complex convolution and complex batch normalization. However, neither of these works explored any complex-valued activation functions, which could have added value to the comparisons. Also, neither of these works experimented with an unrolled network architecture, which is a model-driven approach which also incorporates known MR physics. Unrolled networks are fairly common in MRI reconstruction [4], [10], [11], [17], [18], [21], [32]–[35]. Therefore, it is important to understand the performance of complex-valued layers on an unrolled network architecture. Additionally, these works do not conclusively evaluate whether complex-valued networks perform better than real-valued networks over a large variety of datasets, network architectures, and network size. Finally, neither of these works show results on any improvements in reconstruction of phase details, choosing instead to focus on the reconstructed magnitude images.

In this work, we investigate a comprehensive case for complex-valued CNNs over real-valued CNNs. We experiment with complex-valued convolution and complex-valued activation functions. We investigate the performance of real-valued and complex-valued CNNs for MRI reconstruction for a variety of architectures, including unrolled networks, network size, and datasets to improve image quality and reduce model size for faster training and more tractable models.

## II. METHODS

### A. Complex-Valued Convolution

We begin with our representation of complex numbers within our convolutional neural network. A complex number can be represented by $d = a + ib$, where $a = Re\{d\}$ is the real component and $b = Im\{d\}$ is the imaginary component. The complex conjugate of $d$ is $\bar{d} = a - ib$.

Instead of separating the real and imaginary components of our data and performing real-valued convolution, we perform the complex-valued equivalent. To do so, we convolve a complex filter matrix $W = X + iY$, where $X$ and $Y$ are real-valued filters, with our complex data $d = a + ib$. Using the distributive property of convolution, we can split this convolution into four separate real-valued convolutions:
$$W * d = (X + iY) * (a + ib)$$
$$= (X * a - Y * b) + i(Y * a + X * b).$$

These convolutions can be represented in matrix form by:
$$\begin{bmatrix} Re(W * d) \\ Im(W * d) \end{bmatrix} = \begin{bmatrix} X & -Y \\ Y & X \end{bmatrix} * \begin{bmatrix} a \\ b \end{bmatrix}.$$

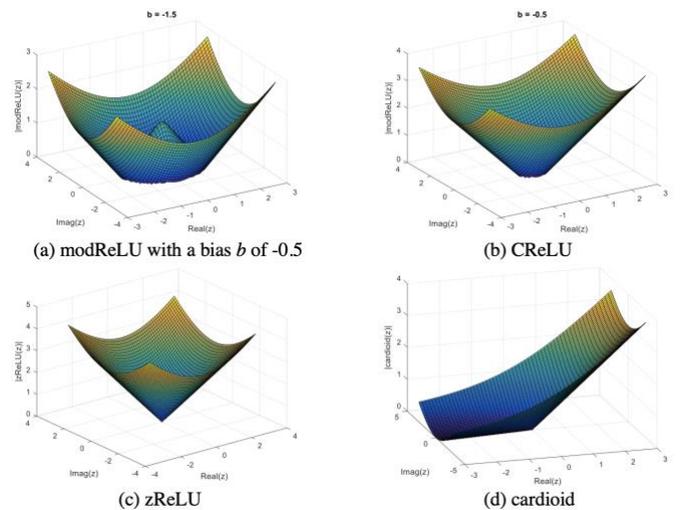

Fig. 1. Surface plots of the four tested complex-valued activation functions.
(a) modReLU with a bias $b$ of -0.5
(b) CReLU
(c) zReLU
(d) cardioid

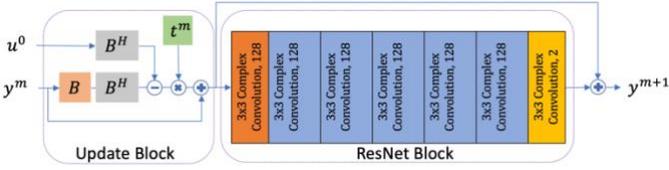

Fig. 2. One iteration of the unrolled network based on the Iterative Shrinkage-Thresholding Algorithm [10], [11]. This consists of an update block, which uses the MRI model to enforce data consistency with the physically measured k-space samples. Then, a residual structure block is used to denoise the input image to produce the output image y_{m+1}. Each convolutional layer except for the last one is followed by a Rectified Linear Unit (ReLU) and a complex-valued activation function (see Section IIB).

For a fair comparison between each of complex-valued and real-valued models, we set the number of feature maps such that each model has the same number of parameters. We explore this later for our model comparisons in our experiments section.

### B. Complex-Valued Activation Functions

There have been numerous activation functions proposed to work with complex numbers. In a standard real-valued CNN, we use the Rectified Linear Unit (ReLU) applied separately to the two channels. In this work, we experiment with modReLU, ℂReLU, zReLU and the cardioid function. The modReLU activation function was originally proposed in the context of unitary Recurrent Neural Networks (RNNs) [36] and tested in deep feed-forward complex-valued networks [22]. The ℂReLU and zReLU functions also have shown promising results in complex-valued convolutional neural networks with applications in speech spectrum prediction and music transcription [22]. The cardioid function has shown promising results for MRI fingerprinting [28]. Surface plots of these activation functions are displayed in Figure 1.

The ReLU function is defined as:
$$ReLU(d) = \begin{cases} d, & \text{if } d \geq 0 \\ 0, & \text{otherwise} \end{cases}.$$

The modReLU function is defined as:
$$modReLU(d) = ReLU(|d| + b)e^{i\theta_d}$$
where $d \in \mathbb{C}$, $b$ is a learnable bias parameter, and $\theta_d$ is the phase of $d$.

The ℂReLU function, which applies separate ReLUs on the real and imaginary components of a complex-valued input and adds them, is defined as:
$$\mathbb{C}ReLU(d) = ReLU(Re\{d\}) + iReLU(Im\{d\}).$$

The zReLU function is defined as:
$$zReLU(d) = \begin{cases} d, & \text{if } \theta_d \in [0, \frac{\pi}{2}] \\ 0, & \text{otherwise} \end{cases}.$$

The cardioid function, which scales the input magnitude but retains the input phase, is defined as:
$$cardioid(d) = \frac{1}{2}d(1 + \cos\theta_d).$$

### C. Network Architecture and Training

MRI reconstruction with CNN has been demonstrated with a variety of network architectures. We chose to use two very different reconstruction networks to compare the performance of real and complex convolution. The first network used is based on an unrolled optimization with deep priors based on the iterative soft-shrinkage algorithm (ISTA) [7], [20]. The unrolled network architecture is shown in Figure 2. This network repeats two different blocks: a data consistency block and a de-noising block. Unrolled networks have been commonly used in state-of-the-art MRI reconstruction due to good performance and having the advantage of reducing the reconstruction solution space. The second network used is based on U-Net, which was originally a network for biomedical image segmentation [14]. U-Net was chosen as another architecture because it is the least similar to an unrolled architecture compared to other networks, but it is also fairly common for MRI reconstruction [7], [15], [37]. The U-Net-based architecture is shown in Figure 3. This network uses contracting and expanding paths to capture information.

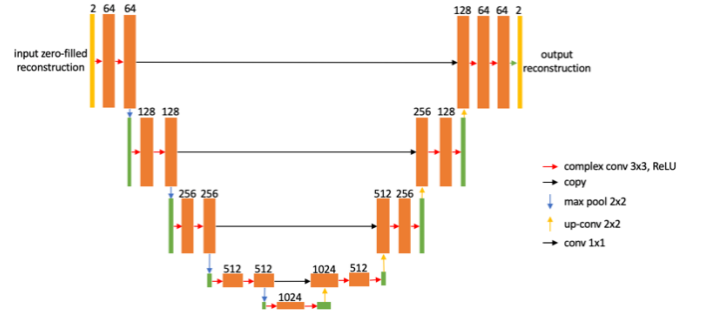

Fig. 3. The second reconstruction network architecture, which is based on the original U-Net for segmentation [14]. Every orange box depicts a multi-channel feature map. The number of channels is denoted on top of each feature map representation. Each arrow denotes a different operation, as depicted by the righthand legend.

The networks were trained with an L1 loss [38] and optimized with the Adam optimizer [39] with $\beta_1 = 0.9$, $\beta_2 = 0.999$, and a learning rate of 0.001. The U-Net was trained with a batch size of 3 and the unrolled network was trained with a batch size of 2. The batch size difference was simply due to different GPU memory limits. All networks were trained for 50,000 iterations. The proposed methods were implemented in Python using Tensorflow. To compute image quality, we evaluated normalized root-mean-square error (NRMSE), peak signal-to-noise ratio (PSNR), and structural similarity index (SSIM) [40] between the reconstructed image and the fully-sampled ground truth. NRMSE and PSNR are evaluated on complex-valued images; however, SSIM is evaluated on magnitude-only images.

### D. Dataset Details

Three sets of data were obtained with Institutional Review Board approval and subject informed consent. First, fully-sampled knee images were acquired using eight coil arrays and a 3D fast spin echo sequence with proton density weighting with fat saturation [41], which we expect to have the least phase variation. Fifteen subjects were used for training and 3 subjects were used for testing. The readout was in the superior/inferior direction, making that direction fully-sampled. Therefore, we subsample in the axial direction. Each subject had a complex-valued volume of size 320x320x256 that was split into axial slices of size 320x256.

The second dataset we used contained body scans which were acquired using 16 coil arrays and an RF-spoiled dual gradient-echo sequence with gadolinium contrast [42], [43].

Here we expect greater phase variation, and the phase should systematically vary between the two echoes depending on the tissue composition. On average, $TE_1$ was 1.1 ms and $TE_2$ was 2.2 ms. The fully-sampled direction was left/right. The dataset was split into 2D coronal slices of 104 patients (21,424 slices) for training, 13 patients (2,646 slices) for validation, and 45 patients (16,098 slices) for testing. Each subject had a complex-valued volume of size 224x220x180 that was split into coronal slices of size 220x180, with each slice considered a separate training example.

Third, fully-sampled 2D phase-contrast cine images were acquired using eight coil arrays in 180 pediatric patients. In this case, the signal phase directly encodes the velocity of blood, which is the clinical goal of the acquisition. In each patient, through-plane velocity was encoded for various vessels of interest including the aorta, pulmonary artery/vein, mesenteric vein, splenic vein, etc. Data acquisition was performed across several 1.5T and 3.0T scanners (Discovery MR 750, GE Healthcare; Waukesha, WI) with a flip angle of 20 degrees, a complex-valued volume of approximate size 192 x 256 x 256, temporal resolution of 50 ms, and venc of 80-500 cm/s based on the clinical scenario. Each cine dataset was split up by cardiac frames and slices, as the network architecture accommodates two-dimensional data.

For each dataset, 72 different variable-density sampling masks were generated using pseudo-random Poisson-disc sampling with acceleration factors ranging from two to nine with a fully-sampled calibration region of $20 \times 20$ in the center of k-space. Sensitivity maps for the data acquisition model were estimated from k-space data in the calibration region using ESPIRiT [44]. The Berkeley Advanced Reconstruction Toolbox (BART) [45] was used to estimate sensitivity maps, generate Poisson-disc sampling masks, and perform a CS reconstruction of these datasets for comparison purposes.

## II. EXPERIMENTS

In the following experiments, we evaluated the effect of number of parameters, network depth, and network architecture on the reconstruction performance of complex-valued and real-valued networks.

In the spirit of reproducible research, we provide a software package in Tensorflow to reproduce the results described in this article. The software package can be downloaded from:
https://github.com/MRSRL/complex-networks-release

### A. Experiment 1: Activation Functions

First, the unrolled network was trained and tested on the knee dataset using first real convolution and then with complex convolution. When real convolution was used, ReLU was applied separately to the real and imaginary channels. When complex convolution was used, the network was additionally trained and tested using the various aforementioned activation functions: modReLU, ℂReLU, zReLU and the cardioid function. The goal of this experiment was to compare the reconstruction performance of real and complex convolution as well as the reconstruction performance of different complex-valued activation functions. The impact of various proposed complex-valued activation functions on reconstruction accuracy was assessed by calculating average NRMSE, PSNR, and SSIM on a test dataset. For all future experiments, we only used ℂReLU as the complex-valued activation function because it performed the best over the other activation functions. The number of iterations and feature maps were fixed for this experiment to 4 and 256, respectively. The real-valued and complex-valued networks were designed to have nearly identical numbers of parameters.

### B. Experiment 2: Complex Convolution and Network Width

Using the knee dataset, we evaluated the impact of the unrolled network's width on the performance of the real versus complex model by fixing the number of iterations to 4 while varying the number of feature maps in each layer. We trained and tested the unrolled network using real and complex convolution over a wide range of network widths with two goals in mind. First, we wanted to see if the performance of the complex-valued model was consistent over many training runs. Second, we wanted to investigate whether the performance of the real-valued and complex-valued models would converge as both models gained more representational capacity.

### C. Experiment 3: Complex Convolution and Network Depth

We investigated if the difference in performance of the real-valued and complex-valued models would converge faster as the number of parameters in each model increased. Using the knee dataset, we varied the depth of the unrolled network by training real and complex-valued networks with 2, 4, 8, and 12 iterations in each layer. The goal of this experiment was to see if the difference in performance of the real-valued and complex-valued models converged as the number of parameters in each model increased.

### D. Experiment 4: U-Net Performance

Additionally, we trained and evaluated the reconstruction performance of two models, one with real convolution, and one with complex convolution, this time using the U-Net architecture. The goal of this experiment was to compare real and complex convolution on an additional architecture to investigate whether the difference in performance of the models was consistent over a variety of network architectures.

### E. Experiment 5: Dual Gradient-Echo Dataset

We then trained and evaluated the unrolled network on the dual gradient-echo dataset for two models, one with real convolution and one with complex convolution. The goal of this experiment was to investigate whether the difference in performance of models with real and complex convolution held up over a variety of datasets. Additionally, the dual-echo dataset has more phase information than the knee dataset, and we believe complex convolution could potentially more accurately represent phase information compared to real convolution. This dataset could be used for fat-water separation, where the phase is important information.

### F. Experiment 6: Phase-Contrast Dataset

We also trained and evaluated the unrolled network on the aforementioned phase-contrast dataset for two models, one with real convolution and one with complex convolution. The goal of this experiment was to investigate whether the difference in performance of models with real and complex convolution held up over a variety of datasets. Additionally, the phase-contrast dataset has more phase information than both the knee and the dual-echo dataset, and we believe complex convolution could potentially more accurately represent phase information compared to real convolution. Phase-contrast data is typically used to view the velocity-encoded images, which is based on the phase of both echoes. Therefore, it is important to accurately reconstruct the phase information of such phase-contrast data. Quantifying flow could be used as a final metric of performance.

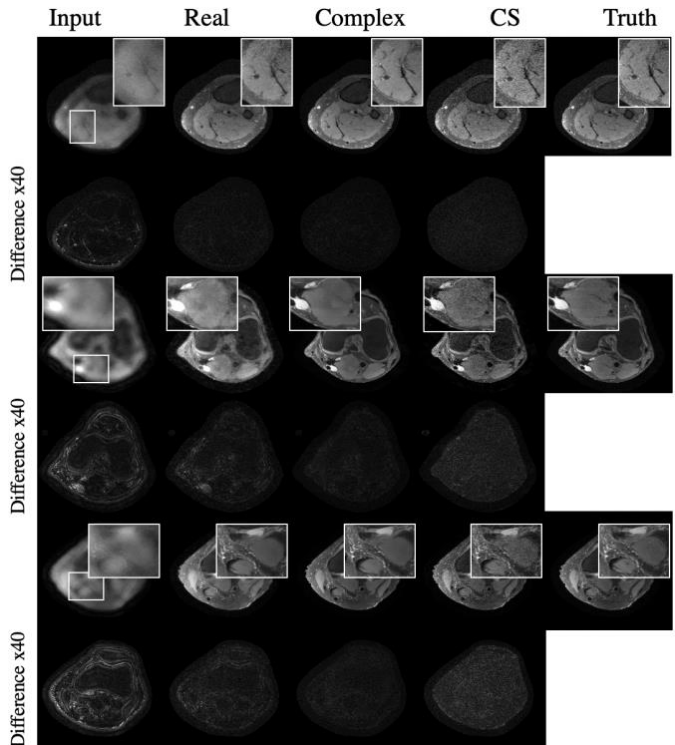

Fig. 4. Representative results from the unrolled network displaying magnitude images, where the left column is the input zero-filled reconstructed image, the second column is the network with real convolution, the third column is the network with complex convolution, the fourth column is the compressed sensing with L1-wavelet regularization reconstruction, and the fifth column is the ground truth reconstruction using all the data. Each row was undersampled by factors of 2.25, 7.40, and 7.37, from top to bottom. The difference maps, scaled by a factor of 40, are displayed under each reconstruction.

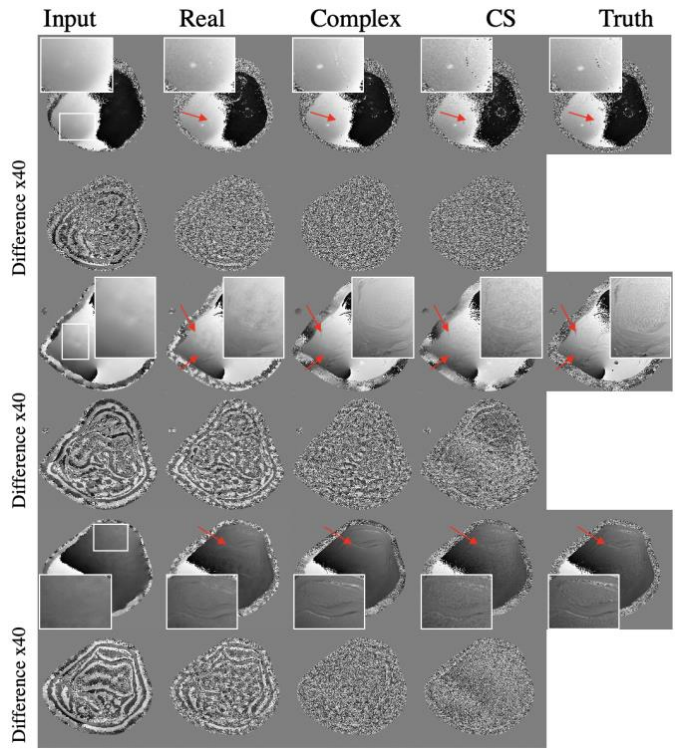

Fig. 5. Representative results from the unrolled network displaying phase images, where the left column is the input zero-filled reconstructed image, the second column is the network with real convolution, the third column is the network with complex convolution, the fourth column is the compressed sensing with L1-wavelet regularization reconstruction, and the fifth column is the ground truth reconstruction. Each row was undersampled by factors of 2.25, 7.40, and 7.37, from top to bottom. The difference maps, magnified by a factor of 40, are displayed under each reconstruction. Red arrows indicate differences in visibility of small details.

## III. RESULTS

### A. Experiment 1: Activation Functions

Representative images from the unrolled models for the spin-echo based knee dataset are displayed in Figures 4 and 5. Here, ReLU was used as the activation function for the real-valued model and ℂReLU was used as the activation function for the complex-valued model. Typically, the complex-valued network produces a reconstruction closer to the ground truth than the real-valued network. Both networks typically outperform CS with L1-wavelet regularization. Most notably, the complex-valued network produces a phase reconstruction with lower average NRMSE, higher average PSNR, and higher average SSIM compared to the real-valued network, as shown in Figure 5. The red arrows in Figure 5 suggest the complex-valued network is able to better preserve and reconstruct phase details compared to the real-valued network.

Comparisons of the various complex-valued activation functions' reconstruction accuracy are shown in Table 1. ℂReLU achieves the best performance overall, with zReLU almost achieving the same performance.

TABLE I
COMPARISON OF IMAGE METRICS ON TEST KNEE DATASETS WITH VARIOUS COMPLEX-VALUED ACTIVATION FUNCTIONS.

| Variable-density subsampling (R = 5.4 ± 0.2) | | | |
|---|---|---|---|
| *Method* | *NRMSE* | *PSNR* | *SSIM* |
| Input Images | 0.72 | 24.73 | 0.764 |
| Real Convolution with ReLU | 0.39 | 30.47 | 0.880 |
| Complex Convolution with: | | | |
| ℂReLU | **0.31** | **32.32** | **0.903** |
| modReLU | 0.38 | 30.67 | 0.867 |
| zReLU | 0.32 | 31.97 | 0.899 |
| cardioid | 0.32 | 31.86 | 0.899 |

### B. Experiment 2: Complex Convolution and Network Width

The performance on a test dataset of the real and complex unrolled models as a function of network width is shown in Figure 6. The number of iterations in each model was fixed at

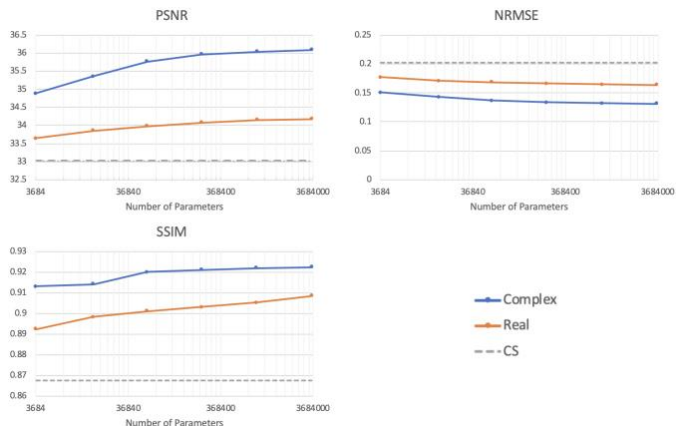

Fig. 6. Performance of the unrolled network as a function of network width on a test dataset. Here, the number of iterations is kept constant at 4, while the number of feature maps is varied for the complex and real networks such that the number of parameters for each evaluation is approximately the same. Compressed sensing does not use feature maps and is plotted for reference. The test PSNR, SSIM, and NRMSE was evaluated for each network.

4 as the number of feature maps was varied, while keeping the total number of parameters for each model approximately the same. When complex-valued convolution was used, the reconstruction performance improved with significantly higher PSNR, lower NRMSE, and higher SSIM. Additionally, the gap in performance between the real and complex models stayed fairly constant as the number of parameters increased.

### C. Experiment 3: Complex Convolution and Network Depth

Similar trends to the results of Experiment 2 can be observed in Figure 7, where the performance of each unrolled model on a test dataset is once again evaluated on the same three image metrics, this time as a function of network depth. Here, the number of feature maps is fixed as the number of iterations was varied, while keeping the total number of parameters for each model approximately the same. The complex-valued model had superior NRMSE, PSNR, and SSIM compared to the real-valued model for all number of iterations.

### D. Experiment 4: U-Net Performance

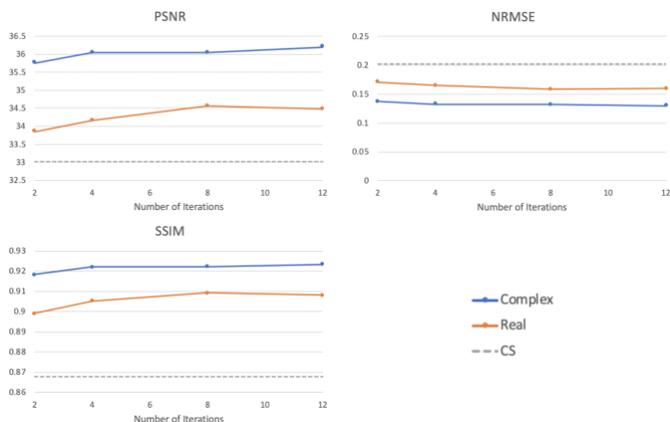

Fig. 7. Performance of the unrolled network as a function of network depth on a test dataset. Here, the number of feature maps is kept constant at 128 and 90 for the complex and real networks, respectively, while the number of iterations is varied for each network. The number of iterations in compressed sensing does not change; however, its performance is plotted for reference. The test PSNR, SSIM, and NRMSE was evaluated for each network.

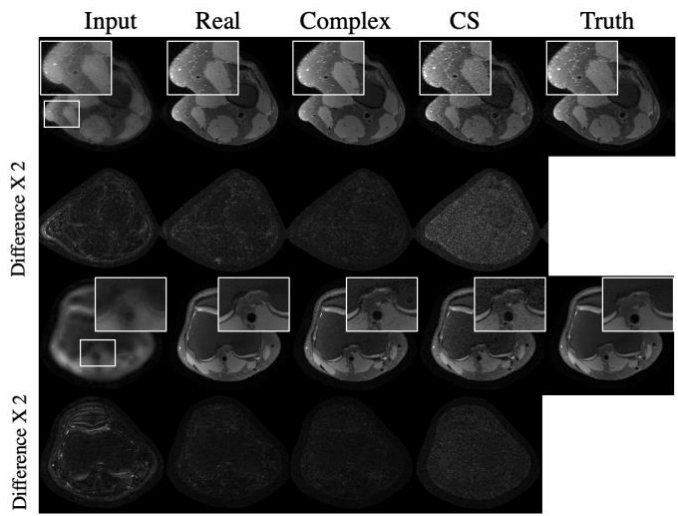

Fig. 8. Representative results from the U-Net displaying magnitude images, where the left column is the input zero-filled reconstructed image, the second column is the network with real convolution, the third column is the network with complex convolution, the fourth column is the compressed sensing with L1-wavelet regularization reconstruction, and the fifth column is the ground truth reconstruction. The top row was undersampled by a factor of 4, and the bottom row was undersampled by a factor of 6. The difference maps, magnified by a factor of 2, are displayed under each reconstruction.

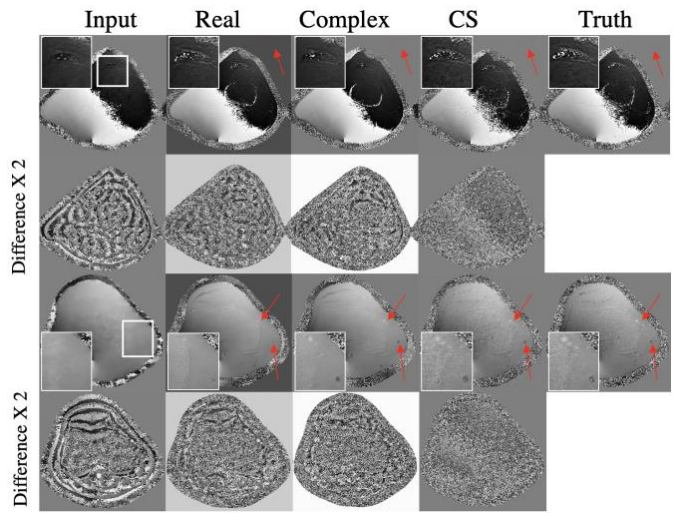

Fig. 9. Representative results from the U-Net displaying phase images, where the left column is the input zero-filled reconstructed image, the second column is the network with real convolution, the third column is the network with complex convolution, the fourth column is the compressed sensing with L1-wavelet regularization reconstruction, and the fifth column is the ground truth reconstruction. The top row was undersampled by a factor of 4, and the bottom row was undersampled by a factor of 6. The difference maps, magnified by a factor of 2, are displayed under each reconstruction. Red arrows indicate differences in visibility of small details.

Representative images from the U-Net are displayed in Figures 8 and 9. Again, the complex-valued network produces a reconstruction closer to the ground truth than the real-valued network. The red arrows indicate differences in the reconstruction of phase details between the various models and CS with L1-wavelet regularization. The real-valued model introduces a phase wrapping error in the background of the phase images which the complex-valued model and CS do not. A comparison of image quality metrics between the real and complex models' performance on a test dataset is summarized in Table 1. When complex-valued convolution was used, the reconstruction performance improved with higher PSNR, lower

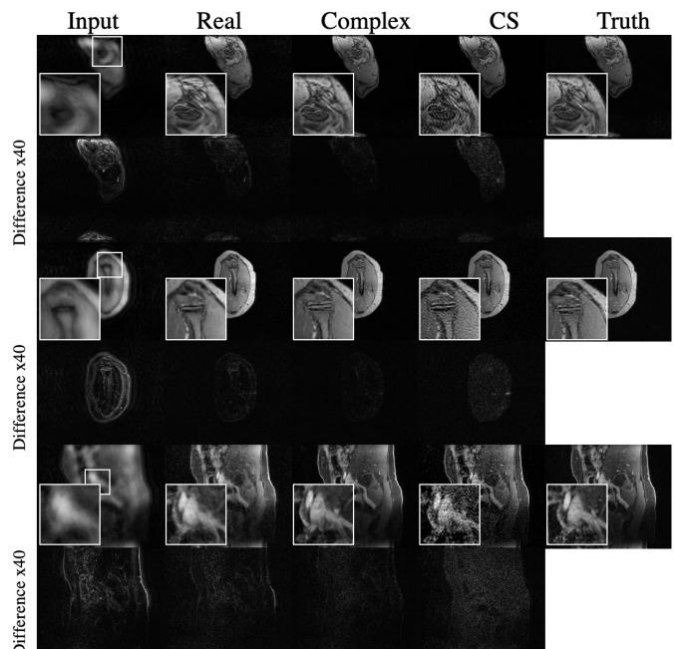

Fig. 10. Representative results from the unrolled network displaying magnitude images from the dual gradient-echo dataset, where the left column is the input zero-filled reconstructed image, the second column is the network with real convolution, the third column is the network with complex convolution, the fourth column is the compressed sensing with L1-wavelet regularization reconstruction, and the fifth column is the ground truth reconstruction. Each row was undersampled by factors of 6, 9, and 4, from top to bottom. The difference maps, magnified by a factor of 40, are displayed under each reconstruction.

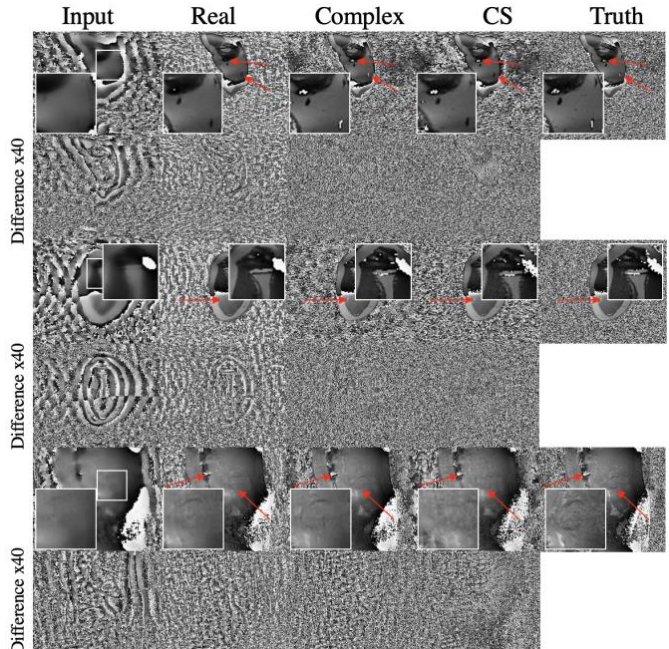

Fig. 11. Representative results from the unrolled network displaying phase images from the dual gradient-echo dataset, where the left column is the input zero-filled reconstructed image, the second column is the network with real convolution, the third column is the network with complex convolution, the fourth column is the compressed sensing with L1-wavelet regularization reconstruction, and the fifth column is the ground truth reconstruction. Each row was undersampled by factors of 6, 9, and 4, from top to bottom. The difference maps, magnified by a factor of 40, are displayed under each reconstruction. Red arrows indicate differences in visibility of small details.

NRMSE, and higher SSIM, despite this model having slightly less parameters than its real-valued counterpart.

### E. Experiment 5: Dual Gradient-Echo Dataset

Representative images from the unrolled network for the full-body dataset are displayed in Figures 10 and 11. The model with complex-valued convolution often produced a much sharper reconstruction. Additionally, we can observe from the difference maps that the complex-valued model produced both a magnitude and phase reconstruction that was visually much more similar to that of the reconstructed ground truth, where the vessels and anatomical structure are much more visible.

### F. Experiment 6: Phase-Contrast Dataset

Representative images from the unrolled network for the phase-contrast dataset are displayed in Figures 12 and 13. In Figure 12, magnitude images from the second echo only are shown. In Figure 13, the velocity-encoded image is shown, which is calculated using both echoes. The model with complex-valued convolution consistently produced sharper reconstructions, where both the magnitude and velocity-encoded images were closer to the ground truth than the model with real-valued convolution and CS with L1-wavelet regularization.

## IV. DISCUSSION

We have introduced the use of complex-valued convolutional layers and complex-valued activation functions to CNNs to significantly improve MRI reconstruction compared to purely real-valued, two-channel CNNs. In a large number of experiments, the complex-valued network achieved superior reconstructions on average compared to a real-valued network with the same number of trainable parameters.

Across different complex-valued activation functions, ℂReLU achieved the best performance over the other, more complicated activation functions. These results suggest a potential for a better performing activation function for complex-valued networks; thus, future work will be directed towards exploring kernel activation functions, which allow the network to learn a trainable function for reconstructions, as described by [28], [46].

When the unrolled network's width and depth were varied over a large range, the complex-valued network still achieved superior image reconstruction metrics with a constant gap in performance compared to the real-valued model, even as the number of parameters greatly increased. Beyond quantitative metrics, the reconstructed images from the complex-valued network better show the details of anatomical structure. One possible explanation for this finding is that the complex-valued model is able to more accurately represent the complex-valued nature of the data compared to the real-valued model, which needs to learn this complex-valued nature. Therefore, the complex-valued model has an inherent advantage regardless of the network size.

The model with complex-valued convolution produced better image reconstructions over the model with real-valued convolution over a variety of architectures, such as an unrolled network and a U-Net, as well as datasets, such as a set of knee images, a set of dual-echo full-body images, and a set of phase-contrast images. We believe there are two possible explanations for this. First, by using complex-valued weights in our model with the complex-valued convolutional layer, we are able to use more feature maps so that the number of parameters is the same as in the model with the real-valued convolutional layer. This enables better reconstruction accuracy. Additionally, because the complex-valued network enforces a structure which preserves the phase of the input data, the reconstructed phase of the complex-valued network is typically much visually closer to the phase of the ground truth compared to the reconstructed phase of the real-valued network. Smaller details in the phase are much more visible in the complex-valued network, as shown by the red arrows in the phase figures.

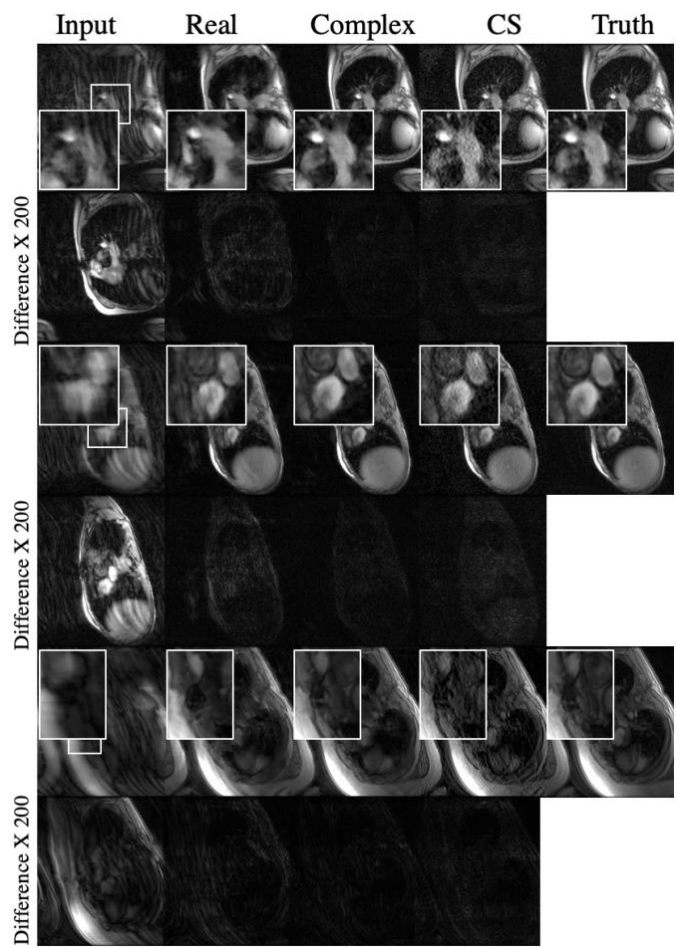

Fig. 12. Representative results from the unrolled network displaying magnitude images from the second echo of the phase-contrast dataset, where the left column is the input zero-filled reconstructed image, the second column is the network with real convolution, the third column is the network with complex convolution, the fourth column is the compressed sensing with L1-wavelet regularization reconstruction, and the fifth column is the ground truth reconstruction. Each row was undersampled by factors of 9, 6, and 4, from top to bottom. The difference maps, magnified by a factor of 200, are displayed under each reconstruction.

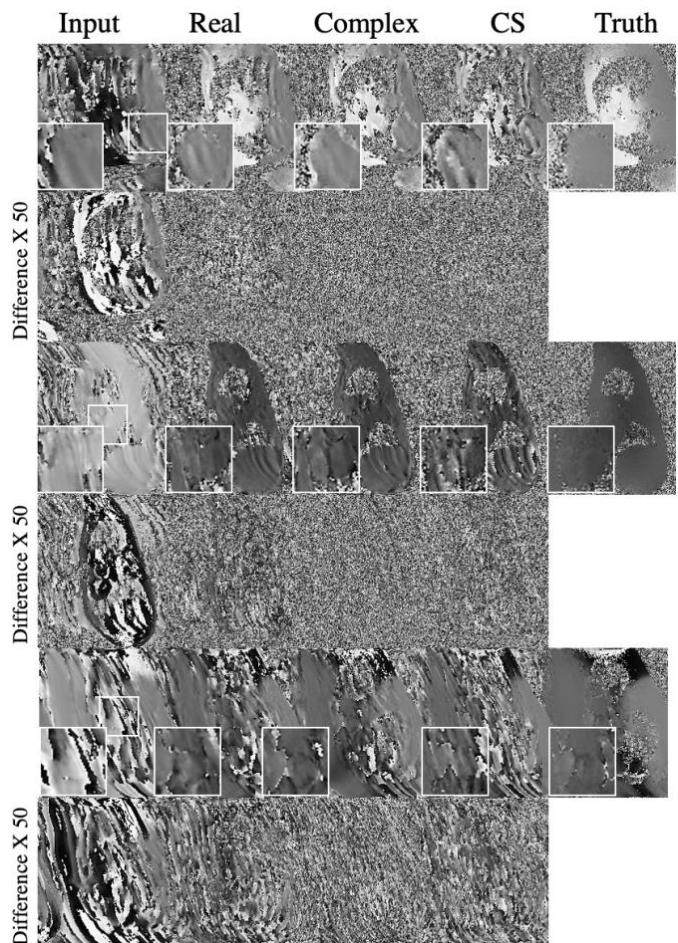

Fig. 13. Representative results from the unrolled network displaying velocity encoded images from the phase-contrast dataset, where the left column is the input zero-filled reconstructed image, the second column is the network with real convolution, the third column is the network with complex convolution, the fourth column is the compressed sensing with L1-wavelet regularization reconstruction, and the fifth column is the ground truth reconstruction. Each row was undersampled by factors of 9, 6, and 4, from top to bottom. The difference maps, magnified by a factor of 50, are displayed under each reconstruction.

It is important to mention that there is value for deep learning approaches over CS in terms of both robustness and quality as well as reconstruction speed. Here, complex networks performed better than CS by some quantitative metrics. However, an advantage of these deep learning reconstructions over CS is greatly reduced reconstruction time.

These methods are extremely generalizable. Here, an unrolled network based on ISTA and a U-Net were used as the two tested network architectures. However, this complex-valued framework can be easily adapted to any other network architecture. Also, the superior performance of complex-valued CNNs can be generalized to many other applications, both in MRI and otherwise. In MRI, this especially has great implications in applications where the reconstructed phase is important, such as in 4D flow, fat-water separation, and QSM. Outside of MRI, complex-valued networks could help deep learning tasks wherever complex numbers are used, including but not limited to ultrasound, optical imaging, radar, speech, and music.

Possible future experiments include adding the complex-valued conjugate of the filter matrix, $W = X - iY$, to the learned feature maps. This could potentially give the network more representative power. This could assist deep learning MRI applications where the physical phenomenon of the complex conjugate is encountered, such as in off-resonance correction. Additionally, complex-valued networks could be extended to quaternion-valued networks with applications in deep learning-based RF pulse design.

## V. Conclusion

In this work, we have explored a variety of complex-valued network architectures with competitive results compared to real-valued architectures. Our work shows that end-to-end complex-valued CNNs provide superior reconstructions compared to real-valued CNNs with the same number of trainable parameters, enabling the potential for reducing MRI scan times by more accurately reconstructing images from subsampled data acquisitions using complex-valued CNNs. Because of superior performance with deep complex-valued networks, we can improve the reconstruction of accelerated MRI scans. We believe the case for complex-valued CNNs can be generalized to other reconstruction architectures, other deep learning MRI applications, and even complex-valued datasets outside of MRI.